\documentclass[final,3p,times]{elsarticle}

\usepackage{amssymb}
\usepackage{amsmath}
\usepackage{multirow}
\usepackage{threeparttable}
\usepackage{longtable}
\usepackage{graphicx,ifpdf}
\usepackage{float}
\usepackage{amsmath}
\usepackage{yhmath,amsthm,amscd,amsfonts,mathrsfs,calrsfs}%
\usepackage{amssymb, upref, color}
\usepackage{fancyhdr,lastpage}%
\usepackage{subfigure}      
\usepackage{booktabs}
\usepackage{geometry}
\usepackage{multirow,setspace,times,amssymb,amsmath,graphicx,color,rotating,subfigure,url}
\usepackage{lineno}
\usepackage{natbib}

\begin{document}

\begin{frontmatter}

\title{Power-law tails in the distribution of order imbalance}
\author[SS,RCE]{Ting Zhang}
\author[RCE,BS]{Gao-Feng Gu}
\author[RCE,BS]{Hai-Chuan Xu}
\author[TJU]{Xiong Xiong}
\author[SZSE]{Wei Chen}
\author[SS,RCE,BS]{Wei-Xing Zhou}

\ead{wxzhou@ecust.edu.cn} %

\address[SS]{Department of Mathematics, East China University of Science and Technology, Shanghai 200237, China}
\address[RCE]{Research Center for Econophysics, East China University of Science and Technology, Shanghai 200237, China}
\address[BS]{Department of Finance, East China University of Science and Technology, Shanghai 200237, China}
\address[TJU]{College of Management and Economics, Tianjin University, Tianjin 300072, China}
\address[SZSE]{Shenzhen Stock Exchange, 5045 Shennan East Road, Shenzhen 518010, China}

\begin{abstract}
  We investigate the probability distribution of order imbalance calculated from the order flow data of 43 Chinese stocks traded on the Shenzhen Stock Exchange. Two definitions of order imbalance are considered based on the order number and the order size. We find that the order imbalance distributions of individual stocks have power-law tails. However, the tail index fluctuates remarkably from stock to stock. We also investigate the distributions of aggregated order imbalance of all stocks at different timescales $\Delta{t}$. We find no clear trend in the tail index with respect $\Delta{t}$. All the analyses suggest that the distributions of order imbalance are asymmetric.
\end{abstract}


\end{frontmatter}

\section{Introduction}
\label{S1:Introduction}

The order imbalance between buyer-initiated and seller-initiated orders is a determinant of stocks price movements. Many scholars have paid attention to investigate the properties of order imbalances. Chordia et al. investigated the determinants and properties of marketable order imbalance on the New York Stock Exchange and the relation among order imbalances, liquidity and daily stock market return \cite{Chordia-Subrahmanyam-2002-JFE}. They found that there exists aggregate contrarian behavior and signed order imbalance is high following market declines and low following market rises. They also found that order imbalance are strongly related to contemporaneous absolute returns after controlling for market volume and market liquidity. Chordia and Subrahmanyam also investigated the relation between estimated order imbalances and individual stock returns \cite{Chordia-Subrahmanyam-2004-JFE}. They demonstrated that individual stock order imbalance is strongly and positively autocorrelated. Further, the relation between lagged imbalances and returns is significantly positive at a one-day horizon. Lee et al. found a similar relationship between stock returns and order imbalance in Taiwan Stock Exchange and they also reported that, even without designated market makers, the exchange is quite effective in absorbing order imbalances \cite{Lee-Liu-Roll-Subrahmanyam-2004-JFQA}. Shenoy and Zhang studied the relation between daily order imbalance and returns in the Chinese stock markets. However, they found that order imbalance cannot predict subsequent returns overall for any subset of stocks, even those with the largest daily price moves. They only found a strong contemporaneous relation between daily order imbalance and returns \cite{Shenoy-Zhang-2007-QREF}.

With the rapid development of China's economy and the increasing capitalization of the Chinese stock markets, much attention is paid to investigate the properties of the emerging Chinese stock market. Both the Shanghai Stock Exchange and the Shenzhen Stock Exchange are order-driven markets containing market orders and limit orders. A market order is submitted to buy or sell a certain amount of shares which results in an immediate transaction. While limit orders that fail to cause an immediate transaction are stored in a queue called limit-order book. All these orders are divided into two groups according to their directions. The amount difference between the buyer-initiated orders and seller-initiated orders is the order imbalance, which can be measured in various ways. In this work, we investigate the probability distribution of order imbalance of Chinese stocks using the order flow data.

The rest of this paper is organized as follows. In section \ref{S1:Database}, we explain the data set adopted and describe in brief the trading rules of the Shenzhen Stock Exchange. Section \ref{S1:PDF} presents the probability distribution of two kinds of order imbalance. Section \ref{S1:Conclusion} summarizes the results.

\section{Data description and trading rules}
\label{S1:Database}

The study is based on the order flow data of 43 liquid stocks traded on the Shenzhen Stock Exchange in 2003. The Shenzhen Stock Exchange was established on December 1, 1990 and started its operation since July 3, 1991. The Exchange is open for trading from Monday to Friday except the public holidays and other dates as announced by the China Securities Regulatory Commission. On each trading day, there is an opening call auction period (9:15 am - 9:25 am), a cooling period (9:25 am - 9:30 am), and two continuous double auction periods (9:30 am-11:30 am and 1:00 pm-3:00 pm). We consider the orders during the continuous double auction in this work.

The database records all the orders of submission and cancelation of 43 stocks, including 32 A-share stocks and 11 B-share stocks in the whole year of 2003. The A-shares are common stocks issued by mainland Chinese companies, subscribed and traded in the Chinese currency Renminbi, listed on mainland Chinese stock exchanges, bought and sold by Chinese nationals and approved foreign investors, and the B-shares are also issued by mainland Chinese companies and listed on mainland Chinese stock exchanges, but subscribed and traded in foreign currency, bought and sold by foreign nationals \cite{Gu-Chen-Zhou-2008a-PA,Gu-Zhou-2009-EPL,Zhou-2012-QF,Zhou-2012-NJP}. The time stamps of the orders are accurate to 0.01 second. The data flow of each stock includes the stock code, trading date, aggressiveness, volume, price, and so on.

\begin{figure}[htb]
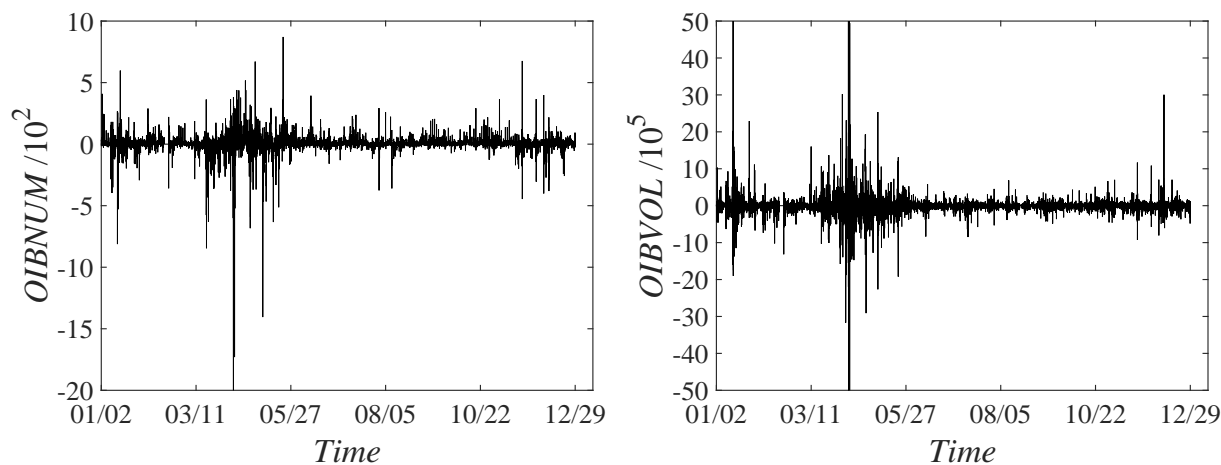

  \centering
  \includegraphics[width=8cm]{Fig_TimeSeries_OIBNUM.eps}
  \includegraphics[width=8cm]{Fig_TimeSeries_OIBVOL.eps}
  \caption{Evolution of the number-based order imbalance OIBNUM and the size-based order imbalance OIBVOL of stock 000001. There are 57840 minutes of the trading day in 2003.}
  \label{Fig:OIB:Data:000001}
\end{figure}

Order imbalance can be defined in various ways \cite{Chordia-Subrahmanyam-2002-JFE}. In this paper, we adopt two definitions based on the number of orders (OIBNUM) and the size of orders (OIBVOL), respectively. The number-based order imbalance OIBNUM is the estimated number of buyer-initiated orders minus seller-initiated trades in one minute and the size-based order imbalance OIBVOL is the estimated size of buyer-initiated orders minus the total size of the seller-initiated orders in one minute. For each stock, we obtain a time series of 57840 data points. In other words, there were 57840 trading minutes in the year of 2003. The two order imbalance time series of stock 000001 are presented in Fig.~\ref{Fig:OIB:Data:000001}. We observe that the order imbalance fluctuates along time. There are outliers with very large order imbalances on 2003/01/14 and 2003/04/16. The stock hit the up-limit price on these two days and the buy limit order books were empty.

The summary statistics of the two order imbalances are presented in Table. \ref{Table:OIB:Summary:Stat}, including the mean $\mu$, the standard deviation $\sigma$, the skewness $S$, the kurtosis $K$, the percentage of positive values $P$\%, and the percentage of negative values $N$\%. For most stocks, the order imbalance is negative and the percentage of positive values is less than the percentage of negative values, which is consistent with the bearish state of the Chinese market in 2003 \cite{Zhou-Sornette-2004a-PA}. The standard deviation of order imbalance varies from stock to stock. We find that OIBNUM tends to be negatively skewed while OIBVOL is more likely to be positively skewed. The kurtoses are significantly greater than 3 (Gaussian distribution), suggesting that the distribution of order imbalance have fat tails.

\setlength\tabcolsep{3pt}
\begin{table}[h!]
  \caption{Summary statistics of order imbalance. OIBNUM is the number-based order imbalance and OIBVOL is the size-based order imbalance. The rows show the stock code, the mean $\mu$, the standard deviation $\sigma$, the skewness $S$, the kurtosis $K$, the percentage of positive values $P$\%, and the percentage of negative values $N$\%.}
  \smallskip
  \label{Table:OIB:Summary:Stat}
  \centering
  \small
  \begin{tabular}{rrrrrrrrrrrrrrr}
  \hline
         & \multicolumn{6}{c}{OIBNUM} && \multicolumn{6}{c}{OIBVOL} \\
\cline{2-7} \cline{9-14}
  Stock  &$\mu$ & $\sigma$ & $S$ & $10^2K$  & $P$\% & $N$\% && $10^2\mu$ &$10^5\sigma$ & $S$ & $10^3K$ & $P$\% & $N$\% \\ \hline
  000001 &     2.55 &    44.55 &    -6.08 &     2.17 &    55.89 &    38.50  &&   -20.91 &     2.07 &    51.10 &     7.07 &    47.61 &    49.99 \\
  000002 &    -1.12 &    22.15 &    -4.38 &     4.09 &    40.70 &    51.84  &&   -38.47 &     0.85 &     5.40 &     0.60 &    43.57 &    54.30 \\
  000009 &    -1.15 &    23.70 &    -5.45 &     1.20 &    48.22 &    45.65  &&   -19.85 &     1.86 &   100.97 &    15.53 &    46.14 &    53.14 \\
  000012 &     0.45 &    12.40 &    -0.84 &     0.39 &    46.07 &    41.51  &&    -8.00 &     0.30 &     1.48 &     0.08 &    46.41 &    49.49 \\
  000016 &    -0.39 &     7.79 &    -2.33 &     0.74 &    42.57 &    44.49  &&    -7.03 &     0.49 &   116.79 &    18.87 &    46.19 &    50.71 \\
  000021 &    -0.60 &    17.96 &    -1.33 &     0.55 &    45.91 &    46.11  &&    -3.03 &     0.73 &    69.75 &     8.58 &    45.76 &    52.42 \\
  000024 &    -0.71 &     6.65 &     8.97 &     7.42 &    35.77 &    45.41  &&    -5.45 &     0.21 &   -10.23 &     0.81 &    42.65 &    48.87 \\
  000027 &    -3.51 &    18.31 &    -5.57 &     1.09 &    35.85 &    53.82  &&   -26.92 &     0.87 &    48.33 &     5.20 &    42.25 &    54.82 \\
  000063 &    -1.92 &    12.26 &    -1.40 &     0.91 &    34.39 &    53.86  &&   -15.01 &     0.42 &     0.24 &     0.26 &    42.19 &    54.18 \\
  000066 &    -0.63 &    14.06 &     7.32 &     8.27 &    44.94 &    44.91  &&    -3.37 &     0.53 &    56.96 &     5.60 &    45.12 &    52.27 \\
  000088 &    -0.55 &     6.94 &    26.37 &    13.87 &    29.07 &    46.67  &&     0.50 &     0.19 &     1.27 &     0.16 &    38.77 &    47.84 \\
  000089 &    -0.11 &     9.10 &     4.13 &     1.42 &    41.91 &    43.32  &&   -21.50 &     0.83 &   -39.70 &     4.31 &    45.62 &    49.54 \\
  000406 &     0.09 &    11.32 &    -1.64 &     0.59 &    48.22 &    41.92  &&   -20.64 &     0.35 &     4.27 &     0.32 &    47.05 &    50.97 \\
  000429 &    -0.41 &     5.53 &    -3.09 &     1.33 &    39.07 &    40.27  &&    -8.27 &     0.22 &     1.09 &     0.22 &    44.71 &    46.43 \\
  000488 &    -0.07 &     4.77 &     1.85 &     0.63 &    39.19 &    40.64  &&   -16.10 &     0.22 &    -3.47 &     0.21 &    40.83 &    49.96 \\
  000539 &     0.16 &     8.71 &    14.02 &     4.82 &    36.32 &    39.75  &&   -11.35 &     0.79 &    48.40 &     6.74 &    39.57 &    47.61 \\
  000541 &    -0.26 &     3.92 &    10.67 &     9.24 &    34.69 &    41.40  &&    -4.77 &     0.14 &    -1.28 &     0.42 &    40.32 &    47.23 \\
  000550 &    -0.78 &    16.92 &     0.12 &     1.39 &    40.48 &    49.10  &&    -4.71 &     0.79 &    60.68 &     6.46 &    44.60 &    52.57 \\
  000581 &    -0.60 &     7.11 &     6.25 &     1.49 &    30.95 &    46.46  &&    -8.70 &     0.25 &    -4.49 &     0.32 &    38.73 &    48.88 \\
  000625 &    -0.46 &    13.86 &     0.83 &     0.48 &    38.42 &    50.00  &&   -14.02 &     0.51 &     3.17 &     0.17 &    43.76 &    51.82 \\
  000709 &    -0.41 &     9.61 &     3.34 &     2.35 &    40.30 &    46.52  &&   -23.53 &     0.50 &     2.42 &     0.32 &    44.05 &    52.40 \\
  000720 &     0.22 &     6.40 &    -0.89 &     0.87 &    41.69 &    30.78  &&     0.23 &     0.14 &     0.55 &     0.11 &    43.46 &    38.05 \\
  000778 &    -0.90 &     6.79 &    -1.96 &     0.50 &    36.36 &    48.71  &&   -10.98 &     0.31 &     6.60 &     0.37 &    41.55 &    53.68 \\
  000800 &    -4.31 &    23.14 &     0.19 &     0.64 &    34.47 &    59.01  &&   -26.96 &     1.36 &    25.11 &     2.26 &    43.66 &    54.59 \\
  000825 &    -1.65 &    18.96 &     3.32 &     1.47 &    36.43 &    54.41  &&   -53.59 &     1.16 &     2.02 &     0.17 &    41.44 &    55.58 \\
  000839 &    -0.83 &    25.85 &    -2.62 &     0.59 &    48.62 &    45.54  &&   -14.93 &     0.74 &    18.95 &     0.83 &    45.89 &    52.92 \\
  000858 &    -0.94 &    11.21 &    -0.85 &     0.43 &    41.11 &    49.00  &&   -18.69 &     0.44 &     1.07 &     0.08 &    44.99 &    53.00 \\
  000898 &    -5.88 &    26.05 &    -2.44 &     0.65 &    31.93 &    60.86  &&   -95.00 &     1.39 &     2.33 &     0.14 &    38.97 &    59.05 \\
  000917 &    -0.29 &     8.50 &     2.41 &     1.16 &    41.61 &    42.94  &&    -6.48 &     0.17 &     9.93 &     0.50 &    44.57 &    49.37 \\
  000932 &    -1.84 &    15.25 &    -2.75 &     0.71 &    41.26 &    48.65  &&   -35.84 &     0.79 &    75.94 &    12.04 &    43.96 &    53.18 \\
  000956 &    -1.19 &    14.85 &    -1.90 &     0.57 &    42.24 &    50.32  &&   -25.11 &     0.53 &     4.84 &     0.32 &    44.64 &    54.01 \\
  000983 &    -1.92 &     9.93 &    -1.75 &     1.55 &    34.28 &    52.55  &&   -14.15 &     0.38 &     3.29 &     0.27 &    42.54 &    53.44 \\
  200002 &    -0.12 &     2.68 &    -0.86 &     0.23 &    32.74 &    35.33  &&    -6.44 &     0.28 &     3.80 &     0.12 &    37.67 &    40.45 \\
  200012 &    -0.10 &     2.75 &    -1.25 &     0.39 &    34.57 &    33.73  &&   -10.38 &     0.25 &    -1.30 &     0.07 &    39.02 &    39.38 \\
  200016 &    -0.05 &     1.99 &    -2.60 &     0.75 &    28.49 &    28.69  &&    -5.68 &     0.16 &   -15.46 &     1.23 &    32.06 &    32.94 \\
  200024 &    -0.13 &     2.25 &    -1.47 &     0.30 &    29.02 &    31.54  &&    -7.02 &     0.15 &    -5.99 &     0.35 &    33.55 &    35.75 \\
  200429 &    -0.08 &     2.64 &    -1.83 &     0.37 &    34.36 &    32.25  &&    -7.67 &     0.26 &    -0.50 &     0.10 &    38.66 &    37.81 \\
  200488 &    -0.23 &     4.94 &    -0.93 &     0.29 &    39.33 &    40.66  &&   -15.33 &     0.46 &     1.48 &     0.11 &    43.96 &    46.45 \\
  200539 &    -0.38 &     4.61 &    -2.40 &     0.37 &    38.61 &    39.87  &&   -11.55 &     0.52 &    -0.15 &     0.05 &    43.94 &    45.09 \\
  200541 &    -0.13 &     1.40 &    -1.17 &     0.32 &    21.40 &    26.74  &&    -1.48 &     0.08 &    -1.00 &     0.13 &    25.29 &    28.88 \\
  200550 &    -0.21 &     4.29 &     1.50 &     1.11 &    35.13 &    37.80  &&    -8.84 &     0.33 &     2.91 &     0.15 &    39.24 &    42.97 \\
  200581 &    -0.26 &     1.86 &    -2.42 &     0.40 &    23.17 &    31.61  &&    -4.89 &     0.13 &    -0.03 &     0.36 &    28.14 &    33.58 \\
  200625 &    -0.66 &     6.72 &    -2.43 &     0.86 &    37.69 &    46.97  &&    -0.14 &     0.82 &     2.38 &     0.37 &    45.77 &    48.47 \\
  \hline
  \end{tabular}
\end{table}

\section{Parametric fits to probability distributions}
\label{S1:PDF}

\subsection{Aggregate sample of all stocks}

\begin{figure}[htb]
  \centering
  \includegraphics[width=8cm]{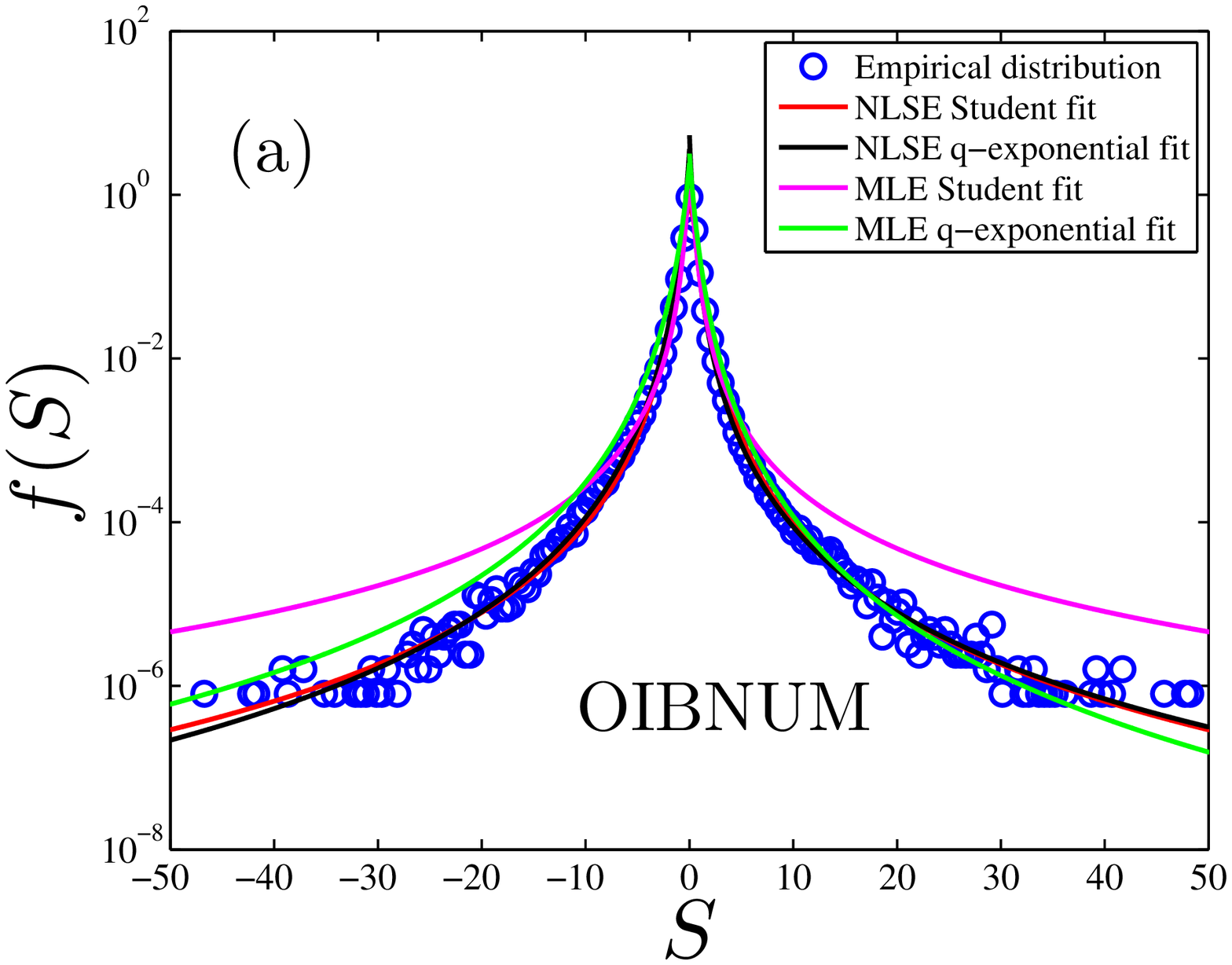}
  \hskip +0.3cm
  \includegraphics[width=8cm]{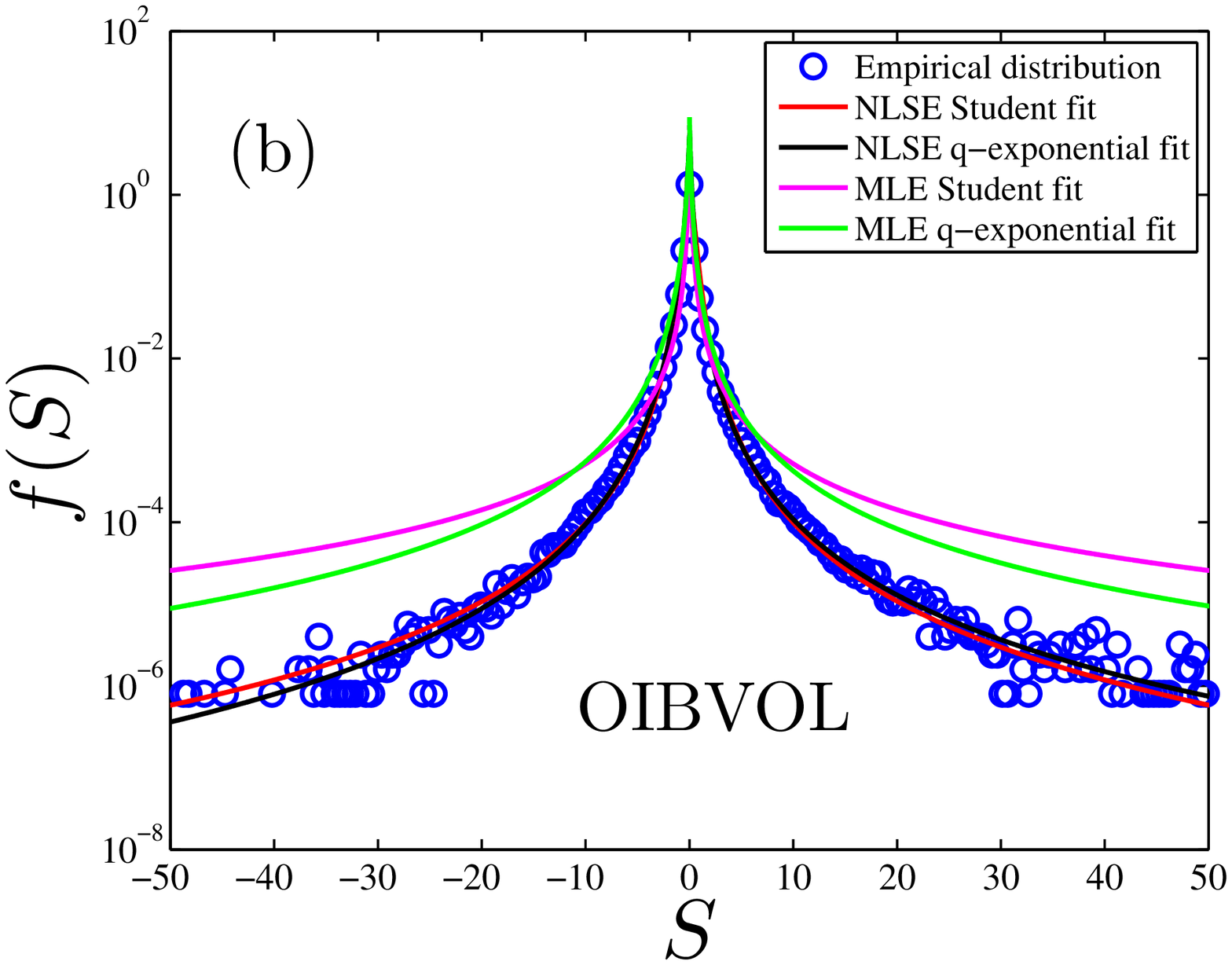}
  \caption{(color online) Empirical distribution of the standardized order imbalance with $\Delta{t}$ = 1 min for all the 43 stocks. (a) Number-based order imbalance OIBNUM. (b) Size-based order imbalance OIBVOL. The red and black line are the nonlinear least-squares fits to the Student and $q$-exponential distributions. The pink and green line are the maximum likelihood fits to the Student and $q$-exponential distributions.}
  \label{Fig:PDF:OIBs:AllStock}
\end{figure}

In order to ensure that the order imbalances of different stocks are comparable, we work on the standardized order imbalance:
\begin{equation}
  S = ({\rm{OIB}}-\mu_{\rm{OIB}})/\sigma_{\rm{OIB}},
  \label{Eq:std:OIBN}
\end{equation}
where OIB can be OIBNUM or OIBVOL, $\mu_{\rm{OIB}}$ is the mean of order imbalance for each stock, and $\sigma_{\rm{OIB}}$ is the standard deviation of order imbalance of each stock. Figure \ref{Fig:PDF:OIBs:AllStock} shows the empirical distributions of the standardized order imbalance aggregated over all the 43 stocks.

Due to the fat-tail feature of the empirical distributions in Fig.~\ref{Fig:PDF:OIBs:AllStock}, we adopt the Student distribution and the $q$-exponential distribution to fit the distributions of the standardized order imbalances $S_{\rm{OIBNUM}}$ and $S_{\rm{OIBVOL}}$, as for the trading volume \cite{Mu-Chen-Kertesz-Zhou-2009-EPJB}. The Student probability density function $f_t(S)$ reads:
\begin{equation}
  f_t(S|\alpha,\mu,L)=\frac{{L}^{\frac{1}{2}}\alpha^\frac{\alpha}{2}}{B\left(\frac{1}{2},\frac{\alpha}{2}\right)}\left[\alpha+L\left(S-\mu\right)^2\right]^{-\frac{\alpha+1}{2}},
  \label{Eq:tpdf}
\end{equation}
where $\alpha$ is the degree of freedom parameter (or tail exponent), $\mu=\langle{S}\rangle$ is the location parameter, $L$ is the scale parameter, and $B(\mathrm{a,b})$ is the Beta function, that means, $B(a,b)=\Gamma(a)\Gamma(b)/\Gamma(a+b)$ with $\Gamma(\cdot)$ being the gamma function. By the definition of $S_{\rm{OIBNUM}}$ and $S_{\rm{OIBVOL}}$ in Eq.~(\ref{Eq:std:OIBN}), we have $\mu=0$. The $q$-exponential probability density $f_q(S)$ is defined as follows \cite{Burr-1942-AMS,Tsallis-1988-JSP,Nadarajah-Kotz-2007-PA}:
\begin{equation}\label{Eq:qpdf}
  f_q(S|\nu,q)=\nu\left[1-(1-q)\nu |S|\right]^{\frac{q}{1-q}},
\end{equation}
where $q$ is usually greater than 1.

When $S$ is sufficiently large, we have
\begin{equation}
  f_t(S)\sim |S|^{-(\alpha+1)}
  \label{Eq:ft:infty}
\end{equation}
for the Student distribution and
\begin{equation}
  f_q(S)\sim |S|^{\frac{q}{1-q}}=|S|^{-\left(1+\frac{1}{q-1}\right)}
  \label{Eq:fq:infty}
\end{equation}
for the $q$-exponential distribution. In other words, both distributions have power-law tails with the tail indices being $\alpha$ for the Student distribution and $1/(q-1)$ for the $q$-exponential distribution. Hence, comparing Eq.~(\ref{Eq:ft:infty}) and Eq.~(\ref{Eq:fq:infty}), we can define
\begin{equation}
  \alpha^{\pm}=1/(q^{\pm}-1)
  \label{Eq:alpha:q}
\end{equation}
and calculate the $\alpha^{\pm}$ values from $q^{\pm}$.

We adopt the maximum likelihood estimator (MLE) to fit the two empirical distributions and illustrate the four fits in Fig.~\ref{Fig:PDF:OIBs:AllStock} for comparison. It is evident that both the Student distribution and the $q$-exponential distribution fit the data well for the standardized order imbalances between $\pm5$, which accounts for $99.49\%$ of the sample for $S_{\rm{OIBNUM}}$, and between $\pm3$, which accounts for $98.61\%$ of the sample for $S_{\rm{OIBVOL}}$. However, there are remarkable discrepancies in the tails since the MLE results are dominated by relatively small order imbalances. In order to better capture the tail behavior of the distributions, we adopt the nonlinear least square estimator (NLSE) for model calibration. The resultant fits are also illustrated in Fig. \ref{Fig:PDF:OIBs:AllStock}.

Table \ref{TB:OIB:Aggregate:TailIndex} reports the estimates of parameters. Since both the $q$-exponential and Student models have two parameters, we can compare quantitatively their performance simply using the root mean square value of fit residuals. According to Table \ref{TB:OIB:Aggregate:TailIndex}, the Student distribution outperforms the $q$-exponential distribution for both $S_{\rm{OIBNUM}}$ and $S_{\rm{OIBVOL}}$, except that the $q$-exponential distribution performs better in the MLE fit of $S_{\rm{OIBNUM}}$. Table \ref{TB:OIB:Aggregate:TailIndex} also shows that the MLE provides better goodness-of-fit. When we compare the tail indices $\alpha$, $\alpha^+$ and $\alpha^-$, we should put more credit on the NLSE results. We find that the results are consistent with each other. The overall tail index is 2.60 for OIBNUM and 2.16 for OIBVOL. Also, the positive and negative tails are asymmetric, as indicated by the skewness in Table \ref{Table:OIB:Summary:Stat}.

\setlength\tabcolsep{5pt}
\begin{table}[h]
\centering
\caption{Estimated tail indices of order imbalances by means of NLSE and MLE. $\chi_t$ and $\chi_q$ stand for the r.m.s. of fit residuals. $\alpha^{\pm}=1/(q^{\pm}-1)$. The $p$ value represents the percentage of stocks preferring the chosen model. }
\smallskip
\label{TB:OIB:Aggregate:TailIndex}
\begin{tabular}{cccccccccccc}
  \hline
      &      & \multicolumn{3}{c}{Student}  && \multicolumn{6}{c}{$q$-exponential}\\
  \cline{3-5} \cline{7-12}
  & & $\alpha$ & $\chi_t$ & $p$  && $q^+$ & $\alpha^+$ & $q^-$ & $\alpha^-$ & $\chi_q$ & $p$ \\
  \hline
  NLSE&$S_{\rm{OIBNUM}}$ & 2.60 & 7.27 & 24/43 && 1.39 & 2.56 & 1.33 & 3.03 & 7.33 & 19/43 \\
      &$S_{\rm{OIBVOL}}$ & 2.19 & 4.93 & 40/43 && 1.47 & 2.13 & 1.40 & 2.50 & 5.03 & 03/43 \\
  \hline
  MLE &$S_{\rm{OIBNUM}}$ & 1.55 & 0.03 &  6/43 && 1.30 & 3.33 & 1.32 & 3.13 & 0.01 & 37/43 \\
      &$S_{\rm{OIBVOL}}$ & 0.86 & 0.02 & 42/43 && 1.72 & 1.39 & 1.62 & 1.61 & 0.11 & 01/43 \\
  \hline
\end{tabular}
\end{table}

\subsection{Individual stocks}

We also adopt the nonlinear least-squares estimator and the maximum likelihood estimator to fit the OIBNUM and OIBVOL samples of each stock to the Student distribution and the $q$-exponential distribution. We obtain the tail index $\alpha$ and $\alpha^{\pm}=1/(q^{\pm}-1)$. By comparing the root of mean square of the fit residuals, we can assess the relative performance of the two distribution models. For $S_{\rm{OIBNUM}}$ using MLE, the $q$-exponential distribution outperforms the Student distribution for more stocks (37/43), and for $S_{\rm{OIBVOL}}$ using MLE, the Student model outperforms the $q$-exponential model for more stocks (42/43). For the fits with NLSE, there are more stocks favoring the Student distribution for both $S_{\rm{OIBNUM}}$ (24/43) and $S_{\rm{OIBVOL}}$ (40/43). These results of relative performance are also shown in the ``$p$'' columns of Table \ref{TB:OIB:Aggregate:TailIndex}.

\begin{figure}[htb]
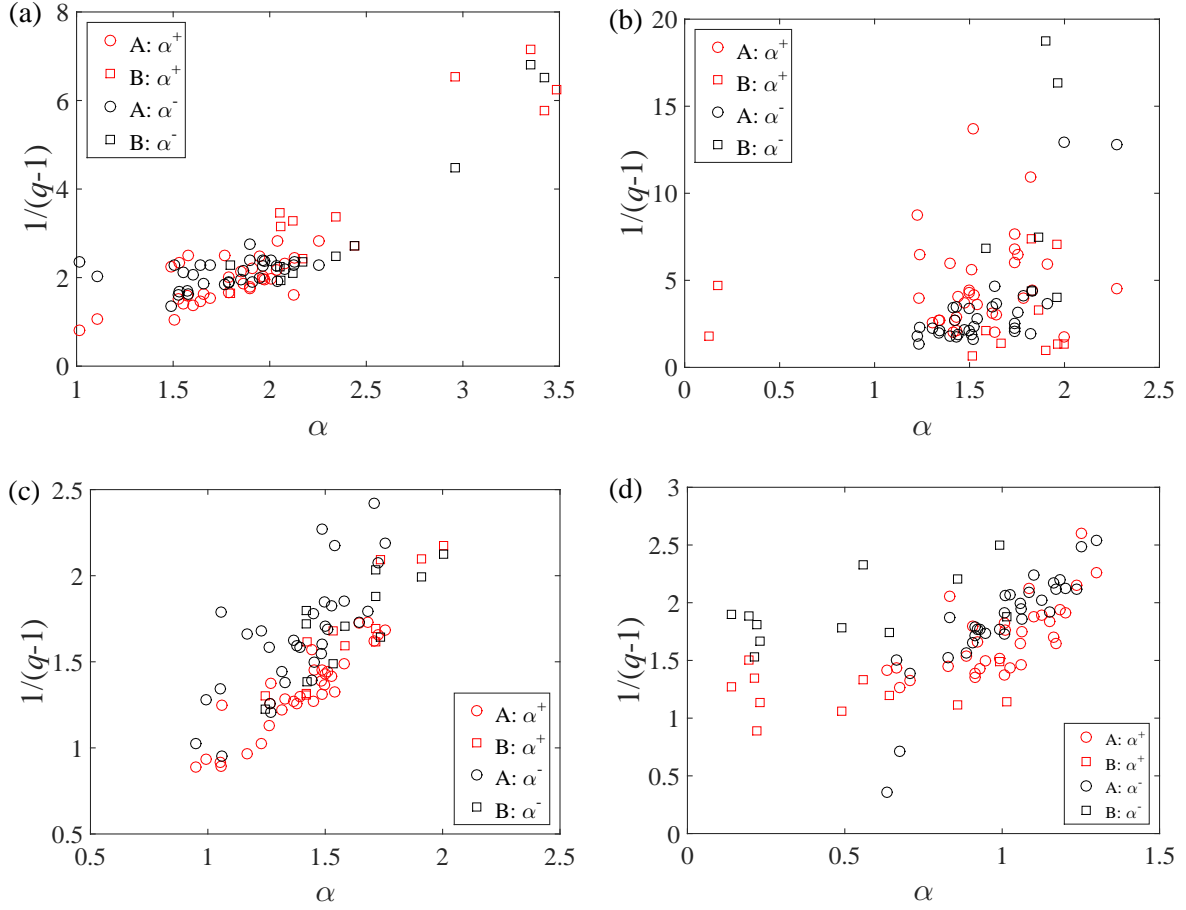

  \centering
  \includegraphics[width=7.5cm]{Fig_TailIndexes_NUM_LSE.eps}\hspace{3mm}
  \includegraphics[width=7.5cm]{Fig_TailIndexes_NUM_MLE.eps}\\
  \vskip 2mm
  \includegraphics[width=7.5cm]{Fig_TailIndexes_VOL_LSE.eps}\hspace{3mm}
  \includegraphics[width=7.5cm]{Fig_TailIndexes_VOL_MLE.eps}
  \caption{(color online) Relationship between tail indices $\alpha$ and $\alpha^{\pm}$ for the number-based order imbalance OIBNUM (a, b) and the size-based order imbalance OIBVOL (c, d), estimated with the nonlinear least-squares estimator (a, c) and the maximum likelihood estimator (b, d). ``A'' and ``B'' in the legends stand for the 32 A-share stocks and 11 B-share stocks.}
  \label{Fig:OIB:TailIndex}
\end{figure}

In Fig.~\ref{Fig:OIB:TailIndex}, we illustrate the scatter plots of tail indices $\alpha$ and $\alpha^{\pm}$ for the number-based order imbalance OIBNUM (upper row) and the size-based order imbalance OIBVOL (lower row), estimated with the nonlinear least-squares estimator (NLSE, left column) and the maximum likelihood estimator (MLE, right column). We find that the tail indices differ from stock to stock. However, there is a positive correlation between $\alpha$ and $\alpha^{\pm}$. The linearity for NLSE plots is better than MLE plots, which is consistent with the fact that the NLSE can better capture the tail behaviors. Indeed, there are negative values of $\alpha^-$ for three B-share stocks and large values of $\alpha^+$ for two B-share stocks in Fig.~\ref{Fig:OIB:TailIndex}(b), which are not shown for better visibility. In Fig.~\ref{Fig:OIB:TailIndex}(a), on average, the OIBNUMs of B-share stocks have larger tail exponents. In Fig.~\ref{Fig:OIB:TailIndex}(c) and (d), the negative tails have larger tail exponents. In Fig.~\ref{Fig:OIB:TailIndex}(d), the OIBVOLs of B-share stocks have smaller tail exponents.

\newpage

\section{Nonparametric determination of tail indices}

\subsection{The case of 1-min order imbalance}

In the preceding section, we have shown that the Student distribution or the $q$-exponential distribution can fit the empirical distributions well and they both have the power-law tails $f(S)\sim S^{-(\beta+1)}$. For large values of $|S|$, the Student density function $f_t(S)$ approaches power-law decay in the tails $f_t(S)\sim|S|^{-(\alpha+1)}$ with a tail index of $\alpha$, the $q$-exponential density function $f_q(S)$ approaches $f_q(S)\sim|S|^{-q/(q-1)}$ with a tail index of $1/(q-1)$.

\begin{figure}[H]
  \centering
  \includegraphics[width=7.5cm]{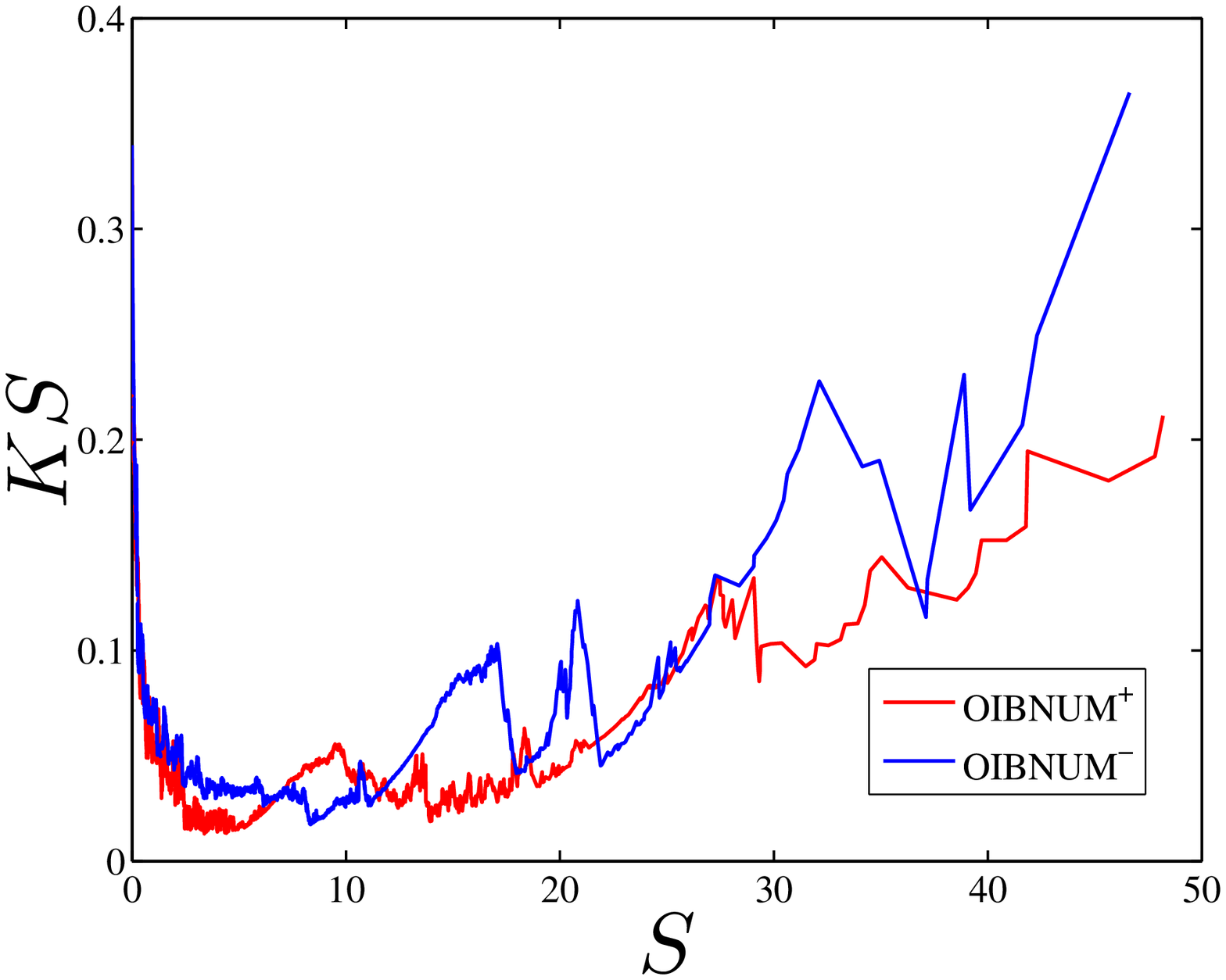}
  \includegraphics[width=7.5cm]{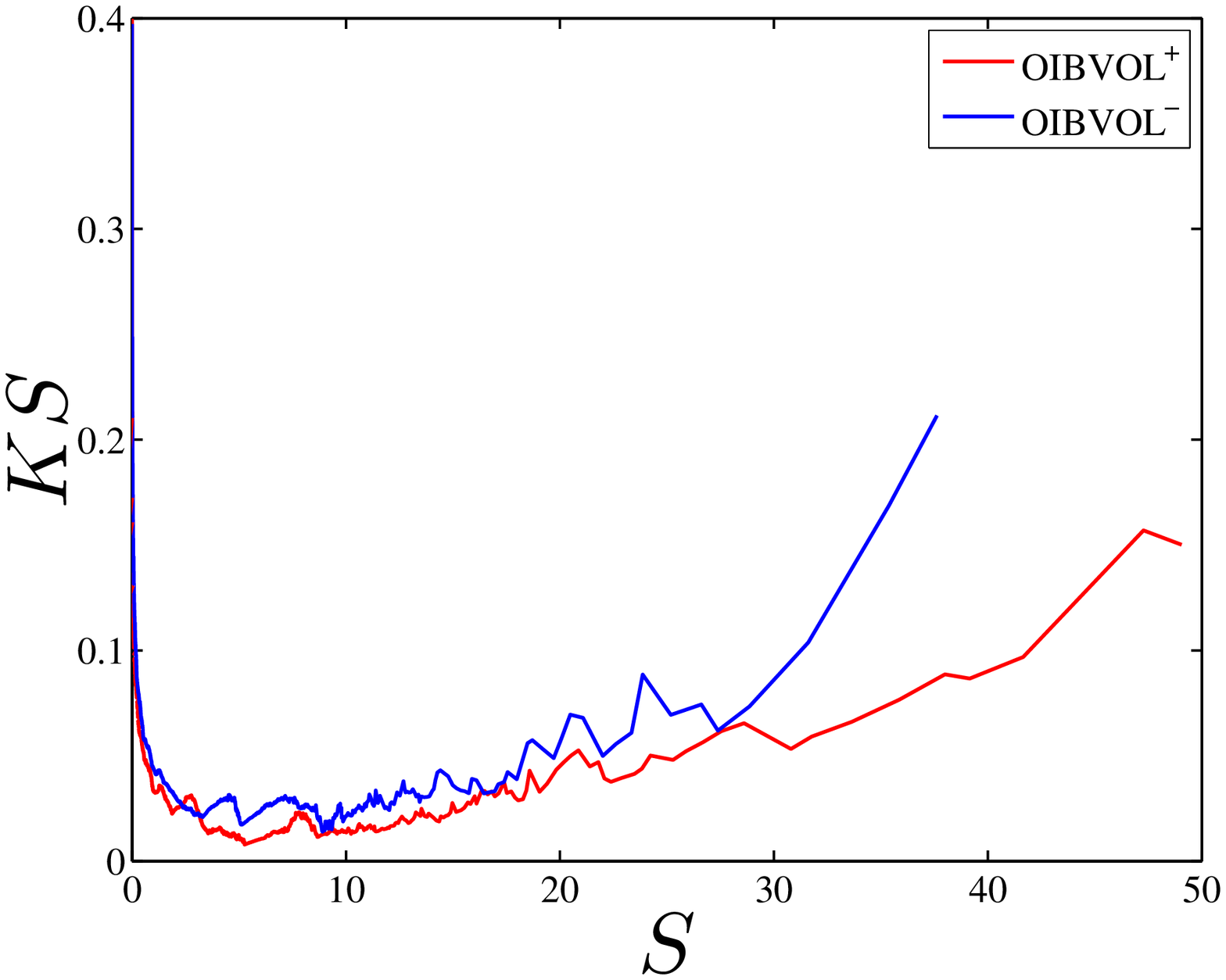}\\
  \vspace{3mm}
  \includegraphics[width=7.5cm]{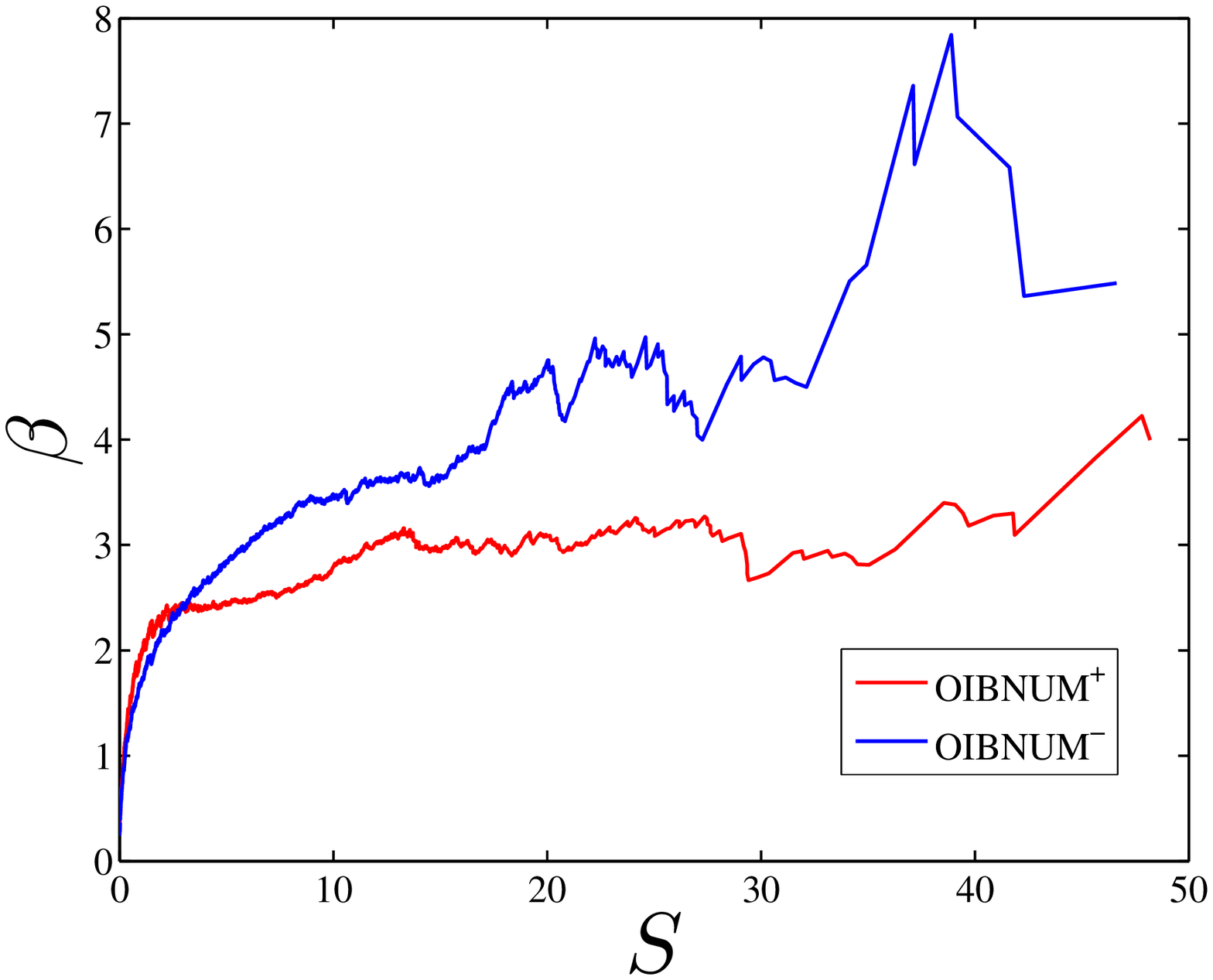}
  \includegraphics[width=7.5cm]{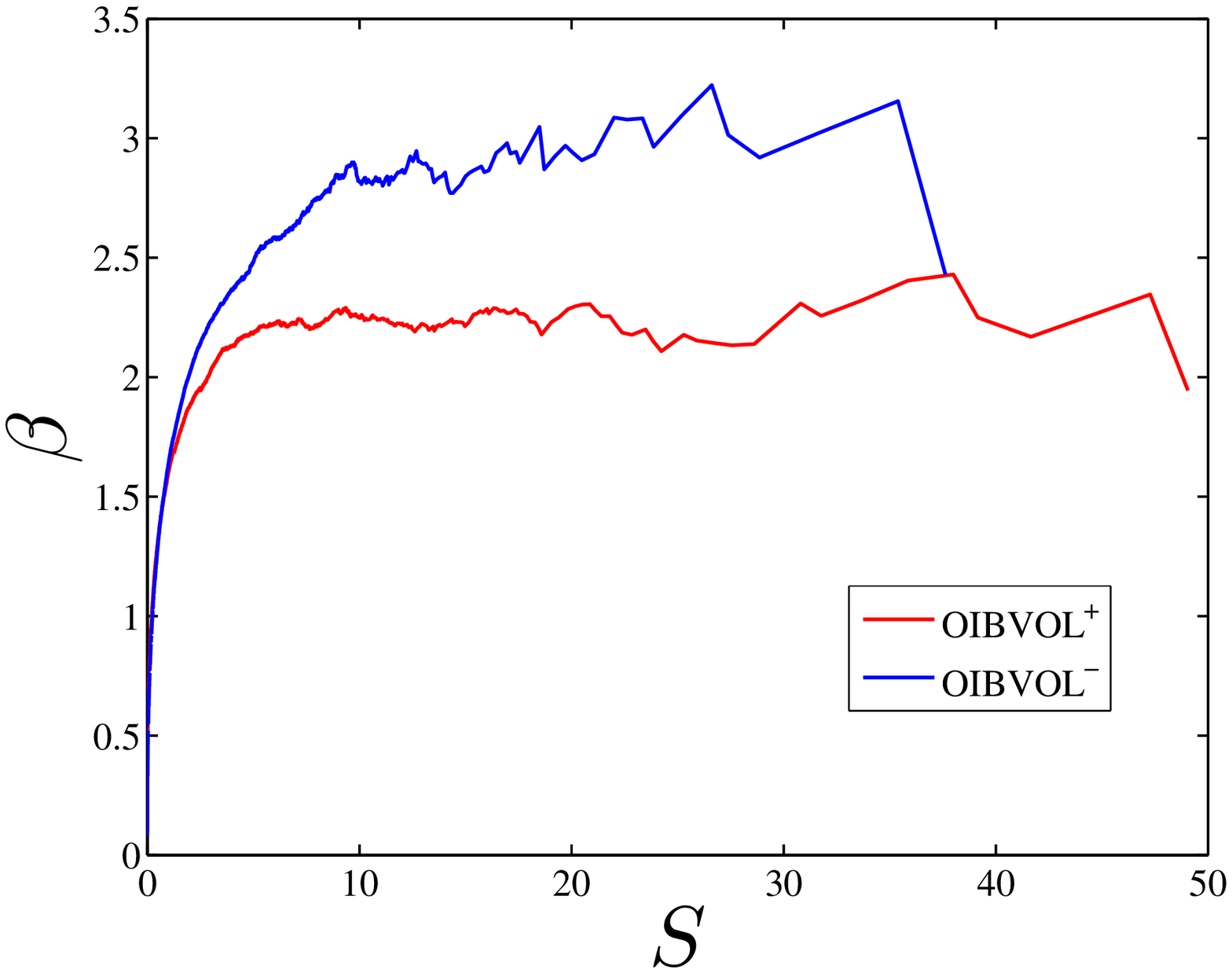}
  \caption{(color online) Determining the cutoff parameter $S_{\min}$ and the tail exponent $\beta$ for the positive and negative tails of the number-based order imbalance OIBNUM and the size-based order imbalance OIBVOL. }
  \label{Fig:KS:OIBN}
\end{figure}

We further investigate the tails of the empirical distribution by using the nonparametric statistical techniques for making accurate parameter estimation of tail indices, based on the maximum likelihood method and the Kolmogorov-Smirnov (KS) statistical test \citep{Clauset-Shalizi-Newman-2009-SIAMR}. The tail index of a sample is estimated based on the data $\left\{S_i:i=1,\cdots,n\right\}$ that are greater than a threshold $S_{\min}$ \citep{Clauset-Shalizi-Newman-2009-SIAMR}:
\begin{equation}
  \beta = n\left[\sum_{i=1}^n\ln\left(\frac{S_i}{S_{\min}}\right)\right]^{-1}.
  \label{Eq:CSN:beta}
\end{equation}
We vary the value of $S$ to obtain the $KS$ statistic and the tail exponent $\beta$. One chooses the threshold value $S_{\min}$ that minimizes the KS statistic and makes the probability distribution of the data above $S_{\min}$ and the fitted power law as similar as possible. In Fig. \ref{Fig:KS:OIBN}(a) and (b), we illustrate dependence of the KS statistics for the positive and negative tails of OIBNUM and OIBVOL as a function of $S$. In Fig.~\ref{Fig:KS:OIBN}(c) and (d), we illustrate the corresponding tail exponents $\beta$ as a function of $S$. It is found that, with the increase of $S$, each $KS$ curve decreases first and then increase, which defines the ``optimal'' value of $S$. We thus obtain that $S_{\min}= 3.38$ and $\beta=2.41$ for the positive tail of $S_{\rm{OIBNUM}}$, $S_{\min}= 8.34$ and $\beta=3.39$ for the negative tail of $S_{\rm{OIBNUM}}$, $S_{\min}= 5.26$ and $\beta=2.21$ for the positive tail of $S_{\rm{OIBVOL}}$, and $S_{\min}= 8.90$ and $\beta=2.85$ for the negative tail of $S_{\rm{OIBVOL}}$.

Furthermore, we adopt the Kolmogorov-Smirnov ($KS$) test and the $CvM$ test \citep{Pearson-Stephens-1962-Bm,Stephens-1964-Bm,Stephens-1970-JRSSB} to check the goodness-of-fit. The Kolmogorov-Smirnov statistic ($KS$) is defined as $KS={\rm{max}}_{S>S_{\min}}(| f(S)-f^*|)$, and the $CvM$ statistic is defined as $C_M^2=n\int_{-\infty}^\infty[f(S)-f^*]^2{\rm{d}}f^*$, where $n$ is the sample size, $f(S)$ is the cumulative distribution of $S$ and $f^*$ is the cumulative distribution of the best power-law fit. The result shows that the two tails of $S_{\rm{OIBNUM}}$ cannot pass the $KS$ test, but they can pass the $CvM$ test. In contrast, both tails of $S_{\rm{OIBVOL}}$ can pass the $KS$ test and the $CvM$ test.

\subsection{Evolving probability distribution of order imbalance}

\begin{figure}[H]
  \centering
  \includegraphics[width=7cm]{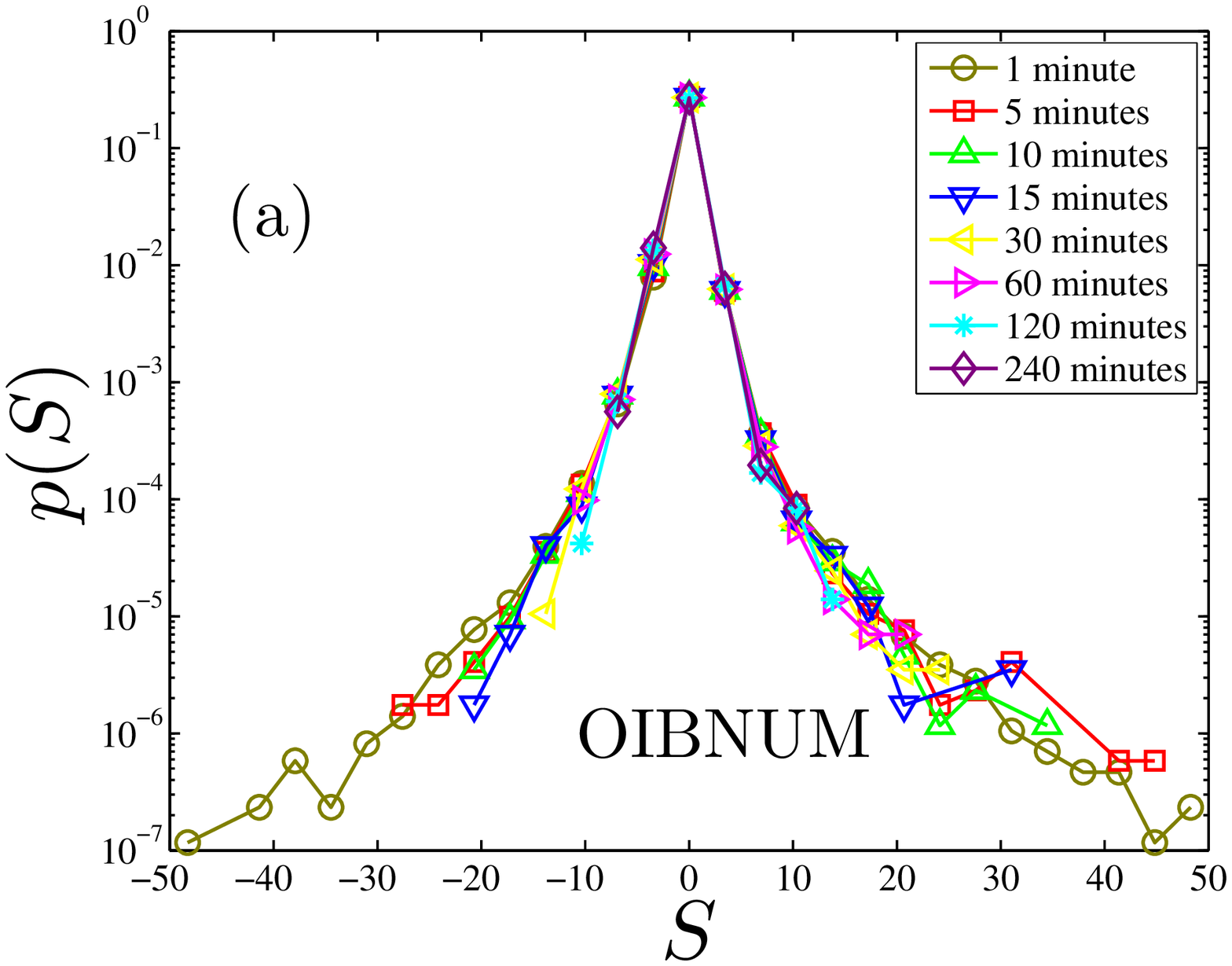}
  \includegraphics[width=7cm]{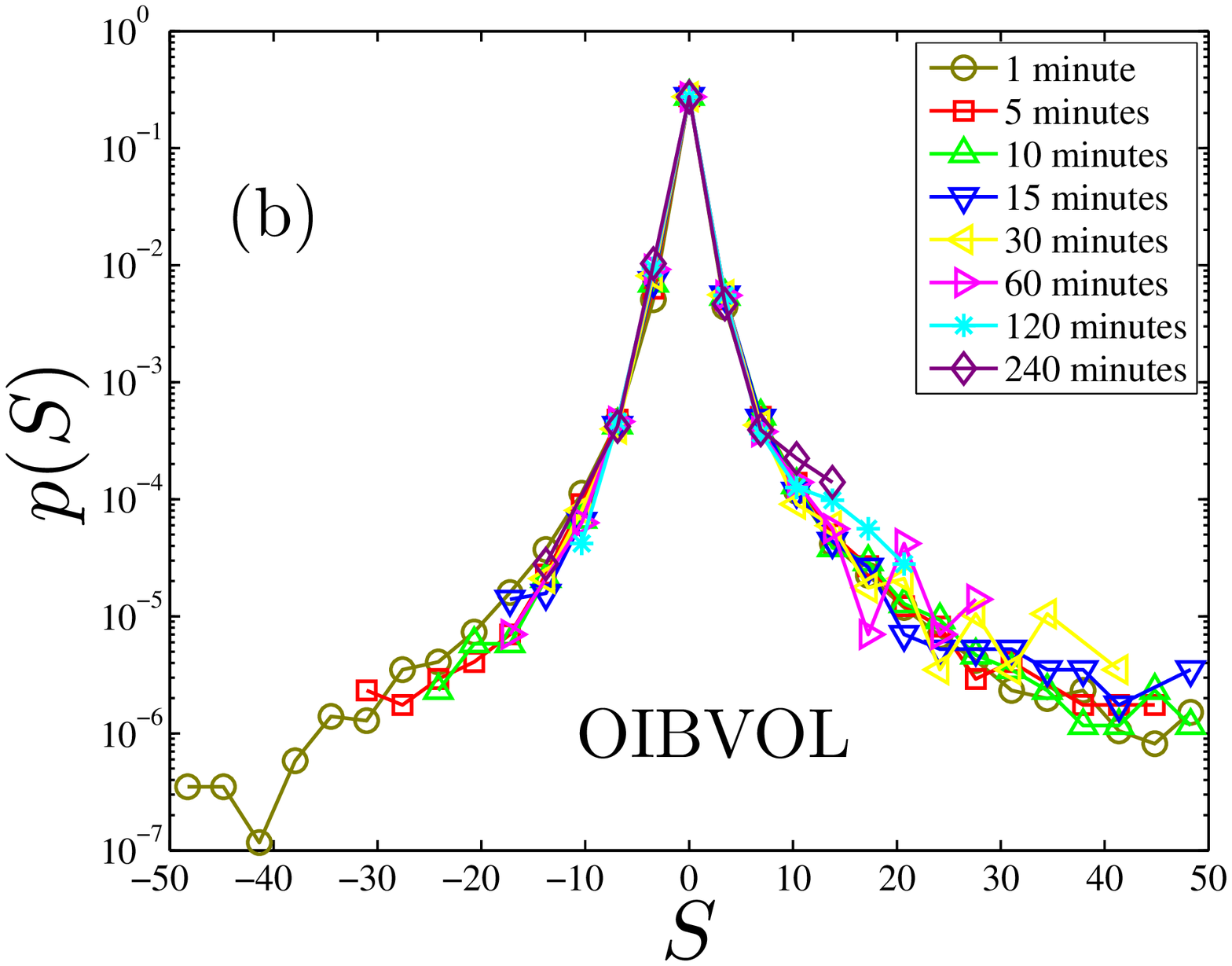}\\
  \caption{(color online) Empirical distributions of aggregated order imbalances OIBNUM (a) and OIBVOL (b) at different timescales $\Delta t$=5, 10, 15, 30, 60, 120 and 240 minutes. }
  \label{Fig:PDF:EVO}
\end{figure}

We now turn to investigate the evolving probability distribution of order imbalance at different timescales $\Delta{t}$. Varying the value of $\Delta{t}$, we are able to compare the distributions at different timescales. Specifically, we compare the distributions for $\Delta{t}$ = 5, 10, 15, 30, 60, 120 and 240 minutes. The empirical distributions $f(S)$ for different timescales $\Delta{t}$ are illustrated in Fig.~\ref{Fig:PDF:EVO}. Interestingly, we find that the distributions at different timescales basically collapse onto a single curve, which is a signal of scaling. However, at the negative tails, we see discrepancies among the distributions to some extent.

In Table~\ref{TB:Evo:PDF}, we present the summary statistics of the order imbalance at different timescales. We find that the number-based order imbalances $S_{\rm{OIBNUM}}$ are left-skewed and the degree of skewness increases with $\Delta{t}$. In contrast, the size-based order imbalances $S_{\rm{OIBVOL}}$ are right-skewed and the skewness decreases with $\Delta{t}$. The kurtosis of each distribution is significantly greater than that of the Gaussian distribution whose kurtosis is 3, indicating a much slower decay of the tails. Moreover, the decreases with increasing $\Delta{t}$ for both $S_{\rm{OIBNUM}}$ and $S_{\rm{OIBVOL}}$.

\begin{table}[H]
  \caption{Characteristic parameters for evolving probability distributions of order imbalances. In the results of the $KS$ test and the $CvM$ test, ``0'' means ``fail to pass the test'' and ``1'' means ``pass the test''. We can see that all these tails can pass the tests.}
  \label{TB:Evo:PDF}
  \smallskip
  \centering
  \begin{tabular}{cccccccccccccccc}
  \hline
   & $\Delta$t& Skewness& Kurtosis& $S_{\min}^+$& ${\beta}^+$& $S_{\min}^-$& ${\beta}^-$& $KS^+$& $KS^-$& $CvM^+$& $CvM^-$\\
  \hline
   $S_{\rm{OIBNUM}}$&  5&  0.00& 79.44 & 2.56& 2.15& 6.27 & 2.59& 1& 1& 1& 1 \\
                  {}& 10& -0.37& 50.55 & 3.70& 2.38& 5.37 & 2.69& 1& 1& 1& 1\\
                  {}& 15& -0.57& 43.92 & 4.00& 2.55& 5.30&  2.65& 1& 1& 1& 1\\
                  {}& 30& -0.65& 30.59 & 2.49& 2.29& 4.39&  2.61& 1& 1& 1& 1\\
                  {}& 60 & -0.69& 24.62 & 2.47& 2.32& 5.35 & 3.01& 1& 1& 1& 1\\
                  {}& 120&-0.82& 16.84 & 1.26 & 1.96& 1.72&  2.81& 1& 1& 1& 1\\
                  {}& 240& -0.83& 12.69 & 1.75& 2.22& 1.09&  1.54& 1& 1& 1& 1\\
  \hline
  $S_{\rm{OIBVOL}}$ &  5& 6.26& $515.33$& 2.47& 1.98& 6.22&  3.15& 1& 1& 1& 1\\
                  {}& 10& 5.82& $367.91$& 4.77& 2.25& 4.26 & 3.13& 1& 1& 1& 1\\
                  {}& 15 &5.12& $290.21$& 1.79& 1.97& 3.93&  3.23& 1& 1& 1& 1\\
                  {}& 30 &4.21& $178.41$& 1.82& 2.05& 3.66&  3.21& 1& 1& 1& 1\\
                  {}& 60 &2.37& $103.24$& 1.44& 2.01& 2.36&  2.81& 1& 1& 1& 1\\
                  {}& 120 &1.93& $63.81$& 1.25& 2.02& 3.08&  3.43& 1& 1& 1& 1\\
                  {}& 240 &1.58& $43.42$& 0.90& 2.09& 3.89&  4.36& 1& 1& 1& 1\\
  \hline
  \end{tabular}
\end{table}

We determine the tail indices using the nonparametric method \cite{Clauset-Shalizi-Newman-2009-SIAMR}. The results are also presented in Table~\ref{TB:Evo:PDF}. We do not find clear evidence of trend in the dependence of $\beta$ against $\Delta{t}$. When the cutoffs $S_{\min}$ of two tails are close to each other, the tail indices are close too. The $KS$ test and the $CvM$ test for positive and negative tails show that all these fits to the tails can pass the tests.

\section{Conclusion}
\label{S1:Conclusion}

Order imbalance is an important variable in the study of order-driven markets. In this paper, we have carried out empirical investigations on the distributions of order imbalance defined in two different ways, using the order flow data of 43 stocks traded on the Shenzhen Stock Exchange within the whole year of 2003.  We found that the Student distribution or the $q$-exponential distribution can fit the empirical distribution well. The nonparametric method \cite{Clauset-Shalizi-Newman-2009-SIAMR} confirmed the presence of power-law tails in the order imbalance distributions. We also investigated the order imbalance distributions at different timescales and found that the order imbalance distributions are asymmetric and have power-law tails.

The asymmetry of the order imbalance distributions that the power-law indices of left tails are greater than those of right tails might relate to the market conditions. The Shenzhen Stock Exchange (SZSE) Component Index was 2689.49 on 2003/01/02 and rose to 3479.80 on 2003/12/31 experiencing a rise of 29.39\%. Note that the stocks investigated in this work were the constituents of the SZSE Component Index. The overall rise of the stocks indicates that there are more buy orders than sell orders, pushing the prices up. It results in more data points of positive imbalance than negative imbalance such that the positive tails are heavier than negative tails.

\section*{Acknowledgement}

We acknowledge financial supports from the National Natural Science Foundation of China (71532009, 71501072, 71571121 and 71671066) and the Fundamental Research Funds for the Central Universities (222201718006).


\bibliographystyle{elsarticle-num}
\bibliography{E:/Papers/Auxiliary/Bibliography}

\end{document}